# The GenAI Catch-22: Use of Generative Artificial Intelligence in Norwegian Newsrooms During the 2025 Parliamentary Election

Mari Reisjå
IT University of Copenhagen
marre@itu.dk

Anders Sundnes Løvlie
IT University of Copenhagen
asun@itu.dk

**Abstract**

The increasing use of Generative Artificial Intelligence (GenAI) in journalism raises concerns about possible detrimental effects both on journalism and its democratic function. We explore these risks through a case study of GenAI in Norwegian Newsrooms during the 2025 parliamentary election campaign. Based on interviews with managers and journalists over a ten-month period, we analyse how ambitious visions fared in the face of technological and practical challenges. We highlight the risk of an *internal threat* stemming from the journalists' own use of AI, contrasting the dominant focus on external disinformation threats. We show how newsroom managers shared sociotechnical imaginaries resulting in unrealistically optimistic beliefs about the capabilities of the technology and the pace of development, leading to plans for audience-facing GenAI services collapsing and giving way to more mundane uses of GenAI tools internally in the newsrooms. Furthermore, we identify a vulnerability in the newsroom's resilience against GenAI influence: a *GenAI Catch-22*. In order to monitor the GenAI tools and prevent errors and undue influence, newsrooms rely on human expertise. But by using GenAI extensively, the newsrooms risk a deterioration of human expertise, preventing them from monitoring the GenAI systems adequately.

# 1. Introduction

Generative Artificial Intelligence (GenAI) has rapidly become a part of everyday life in many newsrooms (Simon 2024; de-Lima-Santos et al. 2025; Dodds, Ngai Yeung, et al. 2026; Dodds, Zamith, et al. 2026). While some view this integration optimistically as a technology-driven advance (Opdahl et al. 2023; Kotliar 2026), others have raised concerns about the risk of detrimental effects on democracy (Nemitz 2018; Kreps and Kriner 2023) as well as journalism (Mahony and Chen 2025; Møller et al. 2025).

Previous research has provided valuable insight into how news actors assess GenAI and which areas of use are most widespread (Dodds, Ngai Yeung, et al. 2026; de-Lima-Santos et al. 2025; Fridman et al. 2025; Amponsah and Atianashie 2024). Nevertheless, there is still limited empirical knowledge about how these technologies are applied in journalistic practice over time, and especially in situations where the democratic function of journalism is at stake. Much public discourse and "hype" focus on grand visions of rapid technological change which can lead both to optimistic views of the benefits the technology may bring about, as well as concern for its detrimental side effects. Meanwhile, there is a risk that such grand visions make us blind to small and gradual changes which are introduced without adequate caution. For instance, many scholars have already noted that use of GenAI may lead to increased requirements for quality assurance, which may raise further concerns for editorial autonomy and ethical integrity (Amponsah and Atianashie 2024; Simon 2024; Shi and Sun 2024; Dodds, Ngai Yeung, et al. 2026).

This study explores the implications of journalistic reliance on GenAI using the Norwegian parliamentary election in 2025 as a *crucial case* (Gerring 2007). A national election campaign is a situation where the democratic role of the news media is crucial, and where routines for editorial control and quality assurance are put to the test. This makes election coverage ideal as a crucial case: If unintended technological influence or errors occur during an election campaign, it is highly likely that the same will also happen during "normal" times when journalists and editors are less on guard.

While trust in the news media have been on decline in many parts of the world, Norwegian news media still enjoy a high level of trust from the population (The Norwegian Media Authority 2025). The major national news providers function as shared arenas for information and debate which gather broad audience sections across party-political divides. Following international concerns about the impact of AI on elections (Jungherr et al. 2026; Stockwell et al. 2024), in 2025 a government-appointed expert group highlighted professional journalistic news media as a key element of society's resilience against AI-enhanced disinformation (Schia et al. 2025). At the same time, all the largest news organisations in Norway were experimenting with how to apply GenAI in their work processes in general, and the election coverage in particular. Thus, while news media were expected to act as a democratic counterweight to AI-related risks, they were simultaneously becoming extensive users of the very same technologies.

We explore the tension between the news media's role as a democratic safeguard against AI risks and their own use of GenAI through a longitudinal qualitative study following journalists and editors in four key Norwegian news organizations: the two main public broadcasters, *NRK* and *TV2*, as well as the two largest tabloid newspapers, *VG* and *Dagbladet*. We trace how ambitious initial GenAI visions collapsed and gave way to more mundane internal uses. Furthermore, we identify a structural vulnerability in newsroom oversight, a *GenAI Catch-22*: While newsrooms depend on human expertise to safeguard against GenAI-induced errors and bias, the use of GenAI leads to a risk of internal de-skilling and competence deterioration, ultimately undermining the human oversight required to monitor GenAI systems effectively.

# 2. Related work

## 2.1. Concerns about GenAI as a threat to democracy

GenAI technologies have received much attention since the launch of ChatGPT in 2022, often described as a "hype" (Kotliar 2026). The AI boom has also caused much concern, and researchers warn that AI can rob us of democratic control (Nemitz 2018; Kreps and Kriner 2023). Such threats may be divided into two main categories: *External* and *internal* threats. Threats stemming from *external* forces using AI have been the subject of much attention, often focused on the role of AI in disinformation and manipulation of elections (Bontridder and Poullet 2021; Stockwell et al. 2024; von Sikorski and Hameleers 2025). The Norwegian expert group mapping risks ahead of the 2025 election similarly focused on external threats, and highlighted the role of journalistic institutions in countering such threats (Schia et al. 2025, 67), in particular through fact-checking and verification, which has had a rising importance within journalism, and which AI-generated content may further increase the need for (Steensen et al. 2024; Kavtaradze and Kalsnes 2024).

*Internal* threats, on the other hand, refers to the risk that participants in a democracy – including journalists – can be influenced through *their own use* of GenAI systems. Such threats are difficult to measure and define, as they rest on the sometimes subtle power that technology can have to influence our actions, thoughts and values (Jungherr and Schroeder 2023; Nandini et al. 2024). There is much concern about how large language models (LLMs) – which form the basis of most contemporary GenAI systems, such as ChatGPT – can perpetuate stereotypes, misunderstandings and derogatory forms of expression from the training material, which may disproportionately affect vulnerable and marginalised groups (Bender et al. 2021; Dodge et al. 2021; Gallegos et al. 2024). The trustworthiness of LLM outputs is also much debated, as LLMs are prone to give convincing-looking but inaccurate outputs, sometimes referred to as "hallucination" (McIntosh et al. 2024; Huang et al. 2025; Zhang et al. 2025). In spite of their unreliability, experimental studies have shown that LLMs may be highly effective in influencing people's opinions (Hackenburg et al. 2025; Salvi et al. 2025; Costello et al. 2026; Steyvers et al. 2025). Studies have also demonstrated a troublesome tendency for people to overtrust outputs from LLMs, even in critical domains such as medicine, finance and military applications (Klingbeil et al. 2024; Holbrook et al. 2024; Shekar et al. 2025). This raises the possibility that use of GenAI may influence news journalists in ways that may occur hidden and gradually and be difficult to detect (Shi and Sun 2024).

## 2.2. GenAI in the newsrooms

GenAI is often presented as a way to increase efficiency at a time when media economies are under pressure, and it is argued that GenAI systems may take over simple tasks and free up journalists for investigative journalism (Moran and Shaikh 2022; Opdahl et al. 2023; Fridman et al. 2025). Empirical research indicates that news organisations mainly use GenAI for support and routine tasks such as transcribing interviews and processing large data sets (Amponsah and Atianashie 2024; Simon 2024). However, while automation may save time, several studies have shown that journalists are increasingly given new tasks related to monitoring, control, and quality assurance of GenAI systems (Sonni et al. 2024; Simon 2024; Møller et al. 2025). Dodds and colleagues (2026) argue that the integration of GenAI technologies in journalistic practices can be described as a process of "controlled change", in which journalists proactively use GenAI in a deliberate way to preserve their professional authority, with GenAI functioning as a supplement rather than a substitute for human expertise.

Several studies discuss concerns about whether demands for efficiency may lead to journalists becoming dependent on GenAI technology to carry out their tasks, and whether the combination of time pressure and GenAI use may increase the risk of errors and omissions (Simon 2024; Dodds, Ngai Yeung, et al. 2026). Several studies argue that there is an urgent need for clear guidelines for the use of GenAI in news production (Amponsah and Atianashie 2024; Mahony and Chen 2025). De-Lima-Santos and colleagues (2025) analysed the AI guidelines of newsrooms in 17 countries, and found that almost all guidelines include a strong requirement that humans must always supervise the technology. Dodds and colleagues (2026) note that constant changes to the AI guidelines create uncertainty among journalists about what is allowed and what is not. Several studies highlight the need for transparency around the use of AI, in order to ensure that the process can be controlled and the risk of errors or hidden influence from the systems is minimized (Shi and Sun 2024; Mahony and Chen 2025; de-Lima-Santos et al. 2025).

# 3. Method

This is a qualitative case study of editorial use of generative AI, focusing on a crucial case of the type Gerring (2007) terms "most likely". Coverage of a national election campaign is such a case because we may assume that during this time, journalists and editors are extra observant of routines and control of GenAI technology given that elections are a critical situation for society and the press.

## 3.1. Sampling and collection of empirical data

The news organisations represented in the study were the four largest in the country in terms of digital readership: The two main public broadcasters, *NRK* and *TV2*, as well as the two largest newspapers, *VG* and *Dagbladet*. The study was designed as a longitudinal investigation, based on interviews over a period of ten months from January to November 2025, in order to capture both the planning period ahead of the election campaign, the coverage running up to the election on 8 September 2025, as well as capturing evaluations and reflections in the aftermath. This has made it possible to compare the news organisations' initial plans and visions with how the technology ended up being used in practice.

The data consists of 13 semi-structured interviews with nine informants. Four of the informants were managers responsible for AI adoption (hereafter: "AI managers") who were each interviewed twice: First in the planning phase in January 2025, and then again after the election was over in October 2025. Furthermore, we had intermittent email correspondence with the managers during the ten months period in order to follow the developments in the newsrooms. The five remaining participants were journalists who were responsible for an important part of the election coverage: the party selection quizzes offered by each news website. These quizzes invite voters to answer a number of questions about key political issues and receive a score indicating which parties that most agrees with their own views. These services are used by a large proportion of the population ahead of each national election. 2025 was the first year in which GenAI tools was used in the development of the quizzes, and the five journalists were interviewed about their practical experiences with using GenAI for this purpose. In the case of NRK two journalists were interviewed, as one of them was in charge of the quiz and the other was tasked with making a "party guide" accompanying the quiz (hereafter referred to as "NRK quiz maker" and "NRK party guide maker"). Interviewing both managers and journalists made it possible to examine differences between strategic ambitions and practical implementation of GenAI.

The study was planned and carried out in line with the ethical standards of our university, which does not require formal approval from the Institutional Review Board for this type of study. Each interviewee signed a consent form. In the initial interviews with managers we made agreements to keep their plans confidential until after the election, in order to allow them to speak freely. We agreed with the interviewees that we would withhold their names but attribute quotes with the name of their organisation and their professional role.

## 3.2. Analytical approach

The interviews were recorded and later transcribed for thematic analysis using a codebook structured around three main themes with up to nine sub-codes each. The first theme was *imaginaries and justifications for GenAI use*, which focused on uncovering prevalent sociotechnical imaginaries (Jasanoff 2015) as revealed by commonalities in statements appearing over time and across the various informants. The second theme was *practical use, control and ethics*, referring to patterns that could provide insight into how generative AI affected the journalistic process and its results, linked to theories about power dynamics, automation and expertise (Bråten 1983; Bainbridge 1983; Klein and Hoffman 1993). The third theme was *organisation, power and vulnerabilities*, which looked at how technological infrastructures affected the news organisations' room for manoeuvre in election coverage.

The interviews from each news organisation were analyzed pairwise, comparing the statements made by the manager and journalist(s) in that organisation and summarising findings for each organisation. Subsequently all the quotes were grouped thematically and compared across organisations and roles, examining common tendencies and differences.

## 3.3. Positionality

While this study was conducted the first author (FA) was on study leave from her position as journalist in one of the organisations studied, *NRK*. The manager interviewed was not her direct manager and had not been involved in any of her previous projects. Agreements were made between the FA and her managers to ensure that the project did not conflict with the employer's principles of loyalty. The FA had also previously worked for another of the organisations in the study, *VG*. Such proximity to informants is not unusual in Scandinavian journalism research, as the professional communities are relatively small. These connections helped the FA gain access to key informants in the newsrooms, establish trust with the informants, and understand nuances in their descriptions of their work practices. At the same time, it was important to make agreements that allowed the informants to trust that sensitive information about future plans would not be shared with competitors. The proximity to the field also entails a risk of bias. The FA has reflected on her position throughout the research process and strived for a transparent analysis of the material, while the second author – who is an associate professor with no connection to any of the informants in this study – has offered a critical external perspective.

# 4. Analysis

In the following we present our analysis of the editorial staff's assessment and handling of GenAI before and during the election campaign, including both their initial plans and visions and how these were realized in the election coverage.

## 4.1. Before the elections: Hope, fear and appetite for risk

Ahead of the elections, in January and February 2025, all four news organisations had high expectations regarding their use of GenAI in the election coverage that autumn. In interviews, the AI managers stated as a matter of course that GenAI would play a crucial role in the coverage: "AI can contribute facts and insights that would not otherwise exist, and that can provide important information to voters and the public debate" (NRK AI manager, 3 February 2025). GenAI was consistently referred to as something positive that would influence and change journalism for the better, described through recurring themes such as streamlining, automation, personalization and strengthening journalism. In the context of the election, the managers believed that GenAI could add nuances, depth, customization, and engagement, and benefit their journalistic mission.

In all four media organisations GenAI was deeply integrated into the journalistic process, and the AI managers highlighted positive results from journalistic use of AI, especially regarding efficiency and data analysis. The systems were used for tasks such as idea generation, background investigations, data analyses, generation of interview questions, text improvement, text generation and generating suggestions for follow-up stories. The Dagbladet manager described with enthusiasm how a GenAI tool had increased efficiency in creating so-called "quotation stories", which refers to simple news stories based mainly or entirely on reporting from other media outlets:

> We have a quotation story machine where you can enter a story from *the Daily Mail,* and then a draft will come back in a matter of seconds. The draft is usually very good. The reporter can then go through and fix it a bit, and post it. This has often produced the most read stories in a day. (...) We have examples of people writing three quotation stories in an evening, while at the same time doing actually important journalistic work" (Dagbladet AI manager, 22 January 2025).

The journalist we interviewed from Dagbladet echoed this view, claiming that he now spent much less time than before on creating quotation stories, and therefore could work more on his own journalistic projects. And there he also often used GenAI systems, to help with the initial research: "AI works well as a bulldozer to create a platform to stand on" (Dagbladet journalist, 30 October 2025). While these tools were still controlled by human journalists, the VG AI manager was convinced that the future would bring GenAI systems that were more proactive, and that the processes no longer had to be initiated by humans: "I imagine that we will have our own VG-AI agents who trawl various information flows and web databases and give us feedback on what we should write about" (VG AI manager, 28 January 2025).

In both NRK and VG, the managers were proud of AI-driven investigative stories, in which AI was "used to uncover very important information that did not exist before" (NRK AI manager, 3 February 2025). All four managers saw it as necessary and almost inevitable to use GenAI in journalistic work, stating that it was a goal in itself to test the technology and increase its use among employees. VG stood out with a radical AI-first policy, stating that journalists had to run any new idea through an AI service before discussing it with a (human) colleague. They regularly measured how many of their journalists used the technology at the same time, and during the election they set a new record with 60 percent of their journalists using AI. The goal was 80 percent. The VG manager spoke at times deterministically, predicting that "as the internet killed the paper newspaper, so will AI kill the internet", and described that they had to explore AI in order not to be "shot or eaten by anyone else" (VG AI manager, 13 October 2025).

While the managers were convinced that they had no choice but to use the technology, they also expressed fear about AI hallucination and bias, and that GenAI could cause errors that could weaken the trust of the users. If they used GenAI in a way that damaged the trust this would amount to "shooting oneself in the foot" (TV2 AI manager, 27 January 2025). The NRK AI manager said that readers tended to become skeptical about the content when they were informed that GenAI had been involved. He said that he was unsure about how they should deal with this, because:

> ...we believe that we can actually get completely new, valuable content by using AI. We see that we can deliver a faster and better offering to the audience by using it. But if it leads to the audience having less trust in what we publish, then it is not necessarily worth it. (NRK AI manager, 3 February 2025)

While these concerns did not deter the managers' intentions to use GenAI, they emphasized that use of GenAI should not come at the expense of quality and editorial control: "Everyone has been given a very clear message that reporters must know the story that they put into AI, and that they must carefully review everything that comes out of the AI" (Dagbladet AI manager, 22 January 2025). The managers emphasized that they trusted the journalists to be able to use the systems critically: "Fortunately, we are in a profession where source criticism should be part of the spinal cord" (Dagbladet AI manager, 13 October 2025). The TV2 AI manager said that he had "confidence that good and critical journalists and news editors will see through" news content produced with the help of GenAI (TV2 AI manager, 2 October 2025). In all four newsrooms extensive training had been carried out and AI guidelines had been established which required that a human should approve AI-generated content before publication, and that any exceptions to this should be clearly marked. However, both the managers from Dagbladet and VG also said that the organisations needed to increase their tolerance of errors caused by GenAI: "There is a phase now where you have to experiment, and some mistakes will happen here and there" (Dagbladet AI manager, 22 January 2025). In TV2 and NRK, which are governed by public broadcasting mandates, the risk appetite was somewhat lower: "It's not okay with two or three errors in every news story if AI is involved" (TV2 AI manager, 2 October 2025).

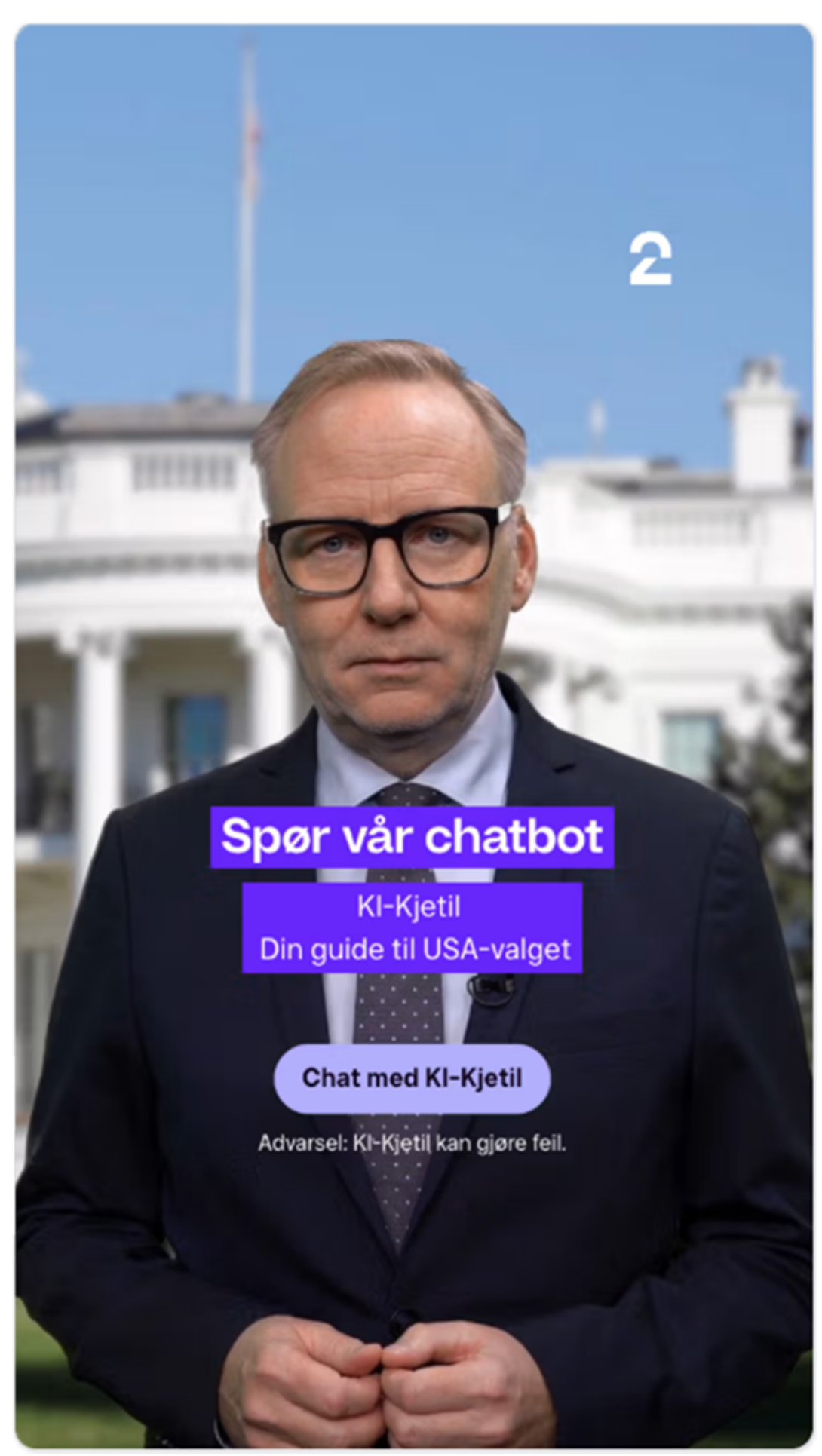

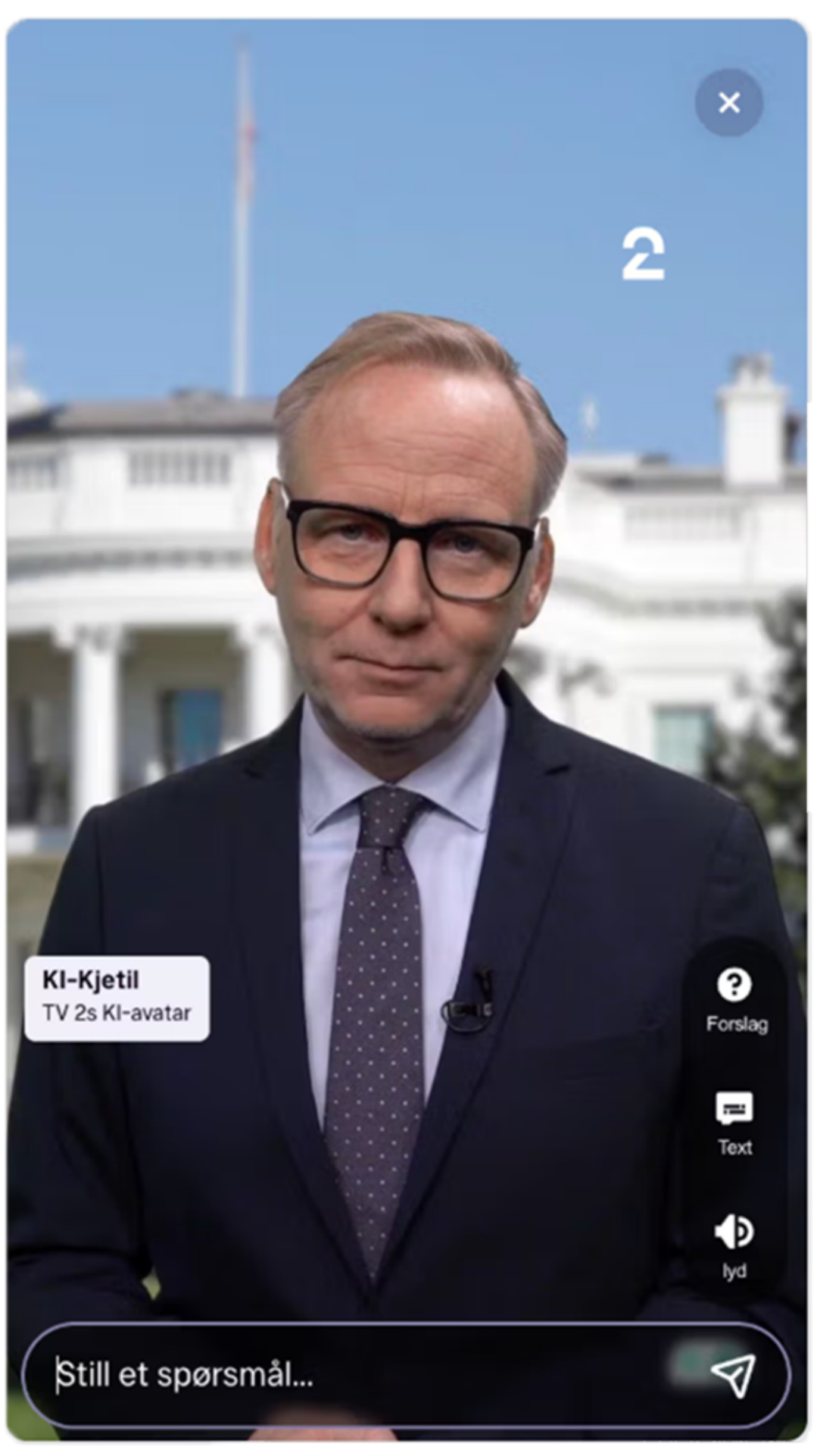

**Figure 1:** Screenshots from "KI-Kjetil" by TV2, a video-based chatbot in which an avatar based on the channel's news anchor Kjetil H. Dale answered questions about the US presidential election in November 2024.

## 4.2. Ambitions shelved in the planning stage

The grand initial ambitions contrasted sharply with what was eventually done in each of the newsrooms during the election campaign. Most of the plans were scaled down or shelved already during early testing.

In January 2025, both TV2, Dagbladet and VG had envisioned that they would offer chatbots which readers could use to get answers about politics, and that these could replace the *party selection quizzes* which have long been a key feature of online election coverage. The VG AI manager said he would be surprised if they did not have a generative, election-specific chatbot. His Dagbladet counterpart stated it was a necessity because he assumed that everyone else would have it. Dagbladet also considered creating chatbots where users could ask questions to avatars of the party leaders. These plans were inspired by concepts successfully tried out during the US presidential election the previous year, when VG had offered a text-based chatbot whereas TV2 had a video chatbot with an avatar based on one of its own news anchors (see Fig. 1). Both these chatbots were quite popular, answering around 100,000 and 70,000 questions while they were active, respectively. NRK was limited by an internal policy that ruled out audience-directed services that used GenAI, but in spite of this they also explored whether GenAI could be used to let people customize the party selection quiz to their own interests and needs, without violating the policy.

Over the course of the spring and summer, all four news organisations shelved their plans for party selection quizzes powered by GenAI. Their justifications for this revealed two shared concerns: First, they had difficulties getting the systems to respond sufficiently accurately. The VG AI manager suggested that the difficulties were caused by the larger number of parties participating in the Norwegian elections, compared to US elections.[1] Furthermore, their demand for accuracy was higher in a domestic context, as readers asked more detailed questions. Second, user tests revealed that readers did not show great interest in election-themed chatbots. NRK's early testing showed that users preferred to indicate their political interests by selecting from a list of options, rather than writing in their own words. TV2 expressed similar insights: "The fact is that you demand more from the user when the user has to actively go in and ask about something, rather than being presented with the curated information you need" (TV2 AI manager, 2 October 2025). VG did launch a general chatbot in the spring of 2025 called "HeiVG", but this was not connected with the election coverage. The chatbot was presented to users as a "non-editorial service", signaling that the news organisation could not fully vouch for the outputs. The chatbot failed to receive the engagement and user numbers they hoped for and was eventually removed in March 2026.

[1] 22 parties were running in the election, of which 9 gained representation in the parliament.

In the initial interviews the managers from the two public broadcasters, NRK and TV2, had envisioned using GenAI to fact-check claims made by politicians in live debates on radio and TV, and each had contracted an external company to develop the technology for it. However, NRK soon had to lower their expectations from imagining that the system could continuously fact-check what the politicians said and show the results to the viewers, to instead considering it as an internal tool that the presenters could lean on during the broadcasts. But even within this more limited scope, the technology failed to deliver. The main problem was that the system was unable to understand which claims were relevant to fact-check:

> There are perhaps 300 claims that appear in one debate, and then there are three of them that we think are journalistically relevant, interesting and important to check. So what we've been working on a lot since then is getting the technology to understand this. We have developed a concept called fact-check worthiness (NRK AI manager, 2 October 2025).

TV2 had envisioned that they could use the technology to develop systems that viewers could interact with on their phones while the TV broadcasts were in progress, but these plans also had to be scrapped: "It was simply not good enough for us to use it for anything" (TV2 AI manager, 2 October 2025).

NRK did carry out an AI project which analysed large amounts of content from the social medium TikTok in order to chart its importance for the election, aiming to uncover misleading narratives or attempts at foreign influence. However, the project ended up being more limited than initially planned, focusing on videos published by parliamentary candidates and analysing who got the most engagement. Technical challenges stood in the way of more complex analysis: "In the same way that it was not possible to go through all of the 10,000 videos manually, it was not possible to go through 10,000 AI-generated fact checks manually" (NRK AI manager, 2 October 2025).

In January and February, the managers had expected that the technology would advance faster than what turned out to be the case when the election campaign started. The post-election interviews revealed internal frustration in all the news organisations that they had not come up with good enough ideas for how they could use the existing technology in a way that benefited them and the audience. They each expressed surprise and relief that none of the competitors had used GenAI to any great extent: "I had thought there would be more of that, but I haven't missed it. It may not be something the audience really needs or demands" (NRK AI manager, 2 October 2025).

Comparing the initial and final interviews, it is clear that the managers changed their perception of what they wanted to use GenAI for. The Dagbladet AI manager had said in January that the parliamentary election was an important area to test and learn about GenAI, while after the election he described use of GenAI in an election campaign as problematic: "Parliamentary elections are quite special. We were a little afraid that we would create services so early on which, what shall we say, trifle with democracy" (Dagbladet AI manager, 13 October 2025). The NRK manager expressed a similar view: "It has become clearer to me what AI is good at and not, and what journalists are good at and not, and that we must find the good match" (NRK AI manager, 2 October 2025).

## 4.3. Use of internal GenAI tools by the journalists

While most of the ideas for public-facing GenAI were cancelled, the news organisations shifted away from thinking about how they could use GenAI to attract attention, to looking at how they could use GenAI as an internal tool. Also here they encountered difficulties achieving the required quality. However, TV2 did implement an internal tool called *Valgvenn*, which had been trained on the programs of the political parties, and was made available to all the reporters. "Valgvenn made some repetitive and heavy tasks easier, such as in the work with creating the party selection quiz. We could spend a little less time on it than we have done before" (TV2 AI manager, 2 October 2025).

As the idea to replace the usual party selection quizzes with chatbots failed, the news organisations reverted to making quizzes in a similar format to previous years. However, for the first time the journalists who created these quizzes integrated internal GenAI tools extensively in their processes (see **Table** 1). These quizzes were an important part of the election coverage, as illustrated by the large number of users that completed one of these quizzes: 2.2 million (VG), 2.9 million (NRK) and 918 000 (TV2), compared to the total of 3.2 million votes in the election. (Dagbladet did not provide numbers of completed quizzes.) This section explores how GenAI was used to create these quizzes, as well as how internal GenAI tools were used in other election coverage.

| | **GenAI systems used** | **Tasks GenAI was used for** |
|---|---|---|
| **NRK (quiz maker)** | • NRK-GPT (internal system based on technology from OpenAI) | • Translate their own text.[2]<br>• Navigate the party programs. |
| **NRK (party guide maker)** | • NRK-GPT (internal system based on technology from OpenAI) | • Condensing and rephrasing of their own text. |
| **TV2** | • Valgvenn (internal system based on technology from Google and OpenAI)<br>• NotebookLM (Google) | • Navigate the party programs.<br>• Generate summaries, which were later rephrased manually. |
| **Dagbladet** | • Claude (Anthropic)<br>• ChatGPT (OpenAI) | • Ask about choice of topics and whether the system could show blind spots they did not cover.<br>• Navigate the party programs.<br>• Extract essence and generate summaries, which were later rephrased manually.<br>• Weighting the parties on a predefined scale. |
| **VG** | • ChatGPT (OpenAI) | • Generate ideas for the design of the party selection quiz.<br>• Generate ideas for questions. At least one question was included without rephrasing.<br>• Simplify and adapt the language to a young audience.<br>• Shorten and simplify a guide for recording video greetings by the party leaders. |

**Table** *1***:** Overview of the main GenAI systems used by the journalists, and what tasks they reported using them for.

[2] Due to public broadcasting requirements NRK needs to produce some content in the minority language Nynorsk, hence the need for translating own text.

All four managers stated that internal GenAI tools had made journalistic work more efficient, which was partly supported by the interviews with the journalists. The TV2 AI manager highlighted that they had spent less time than usual on making the quiz, and the Dagbladet journalist explained that he got help from GenAI to go through the party programs: "I have read most of the party programs, but I don't know every word by heart. So I used our AI tools a lot to find out how the parties stand on our questions" (Dagbladet journalist, 30 October 2025). He felt that using GenAI saved time and explained that he assessed quality by using two external GenAI systems at the same time, comparing the suggestions he received and double-checking the facts where the answers were not consistent.

This enthusiasm for increased efficiency seems to have been coupled with an expectation that GenAI tools could replace some of the need for journalistic expertise. Neither of the informants from VG, Dagbladet and TV2 had any experience making the party selection quiz before, nor did they have much experience with political journalism. At TV2, the task was given to a student intern who had political experience outside of journalism. She understood that she was given the task because the more experienced political journalists were prioritized for other tasks. She received training in how to assess the GenAI outputs critically and found that the two tools she had at her disposal worked relatively well: The internal Valgvenn service and Google's NotebookLM. She liked the latter best, because it provided clear source references to the party programs. Nevertheless, she admitted that there were errors in the quiz that stemmed from her use of GenAI:

> Whenever something looked wrong to me, I asked follow-up questions. But if it didn't, then I didn't consistently ask the follow-up question. So in the end we had four or five errors in the party selection quiz that was on me because I hadn't checked things properly, or that it was an error of definition. (TV2 journalist, 12 November 2025)

She would have preferred to do the work manually if she had the time. She told us about a situation where the GenAI tool introduced an error which she failed to catch due to her lack of expertise: One of the questions generated by GenAI included a politically charged phrase used by right-wing politicians, making the question politically slanted.

> When we sent it [the quiz] around for feedback, an economics journalist said that "working capital" is a controversial concept that those who oppose the wealth tax use as an argument. So I took it out. I'm not very good at economic policy, and these are typical things the AI can overlook. (TV2 journalist, 12 November 2025)

At NRK, the journalist in charge of the quiz had planned to use GenAI extensively, but eventually abandoned the GenAI tool because he discovered that the system produced errors, some of which were serious. NRK had decided early on to expand the party selection quiz compared to previous years, including more parties and more questions, because they believed GenAI would make the work more efficient. When it turned out that they did not get the help they envisioned from the technology, the journalist was left with "an insane amount of work" (NRK quiz maker, 30 September 2025). Thus the most experienced quiz maker interviewed in this study ended up using GenAI the least, as he found it more time-consuming to correct the LLM's suggestions than to perform the tasks himself.

## 4.4. Reduced labelling of GenAI use

At the start of this study in January 2025, all four news organisations had policies requiring that content which had been created with help from GenAI was clearly labelled as such. However, during the time period of our data collection labelling of GenAI use generally became less clear and consistent. For instance, VG changed its AI guidelines from the strict rule that "AI-generated content must be labelled clearly and in a way that is understandable" to a somewhat more vague policy that "VG must provide clear and understandable information about how artificial intelligence is included in journalistic work" (VG 2025). The informants justified these changes by the fact that the technology was used in so many processes that it became too complicated to inform about it always.

> Previously we labelled the stories that were made with the help of AI. We removed that six months ago, because now AI is involved in so many parts of the writing process. (...) So we have removed that label, like most other media outlets. (...) The reader probably also expects it to be used in most cases. (Dagbladet AI manager, 13 October 2025)

The VG AI manager emphasized that principles regarding source criticism apply regardless of which tools are used, and pointed out that they do not normally inform that they had used other types of software such as Microsoft Excel either. Instead, at the time of the election campaign both Dagbladet and VG included links at the bottom of all their articles to general information about AI use in the newsroom. Neither NRK nor TV2 included such links in their articles, but they had public guidelines regarding AI use which could be found by searching for them.

None of the party selection quizzes were labelled with AI use, in spite of the extensive use of GenAI among most of the journalists making them. The journalists in our study varied in their opinions on whether or not the readers should have been informed about their GenAI use. The Dagbladet journalist referred to GenAI as a tool on a par with using the phone: "I don't think we need to say specifically that we have used AI in some project or other. We do that in the same way that we use Google and call people" (Dagbladet journalist, 30 October 2025). The TV2 journalist, on the other hand, thought that they could have been more open about her use of GenAI, and speculated that clearer labelling would strengthen trust in the media in the long run: "What I think weakens trust now is that you don't know what is AI and what is not, both in the media and in society in general. You lose confidence that you can assess for yourself what is AI" (TV2 journalist, 12 November 2025).

The informants mostly dismissed the possibility that GenAI had had an impact on their election coverage. The journalists from NRK, VG and TV2 were convinced that their content would not have been any different if they had not used GenAI: "No, because I used the tools as a kind of complicated search engine that helped me streamline my work. It didn't really have any influence on the result, or how I chose to weight the parties and how I interpreted the answers" (TV2 journalist, 12 November 2025). Only the Dagbladet journalist indicated a possibility that the GenAI systems could have had an impact he was not aware of: "I don't know, that's the honest answer. I don't know how it affected the result other than that it was a very good help to get a result" (Dagbladet journalist, 30 October 2025). It is also worth noting that the informants had little knowledge of how colleagues they collaborated with used GenAI in their processes. The Dagbladet journalist pointed out that while colleagues had checked the quiz before publication, he did not know whether they had used GenAI in their process – raising the possibility that a GenAI tool may have been used for quality assurance of content that had been created by the same tool in the first place.

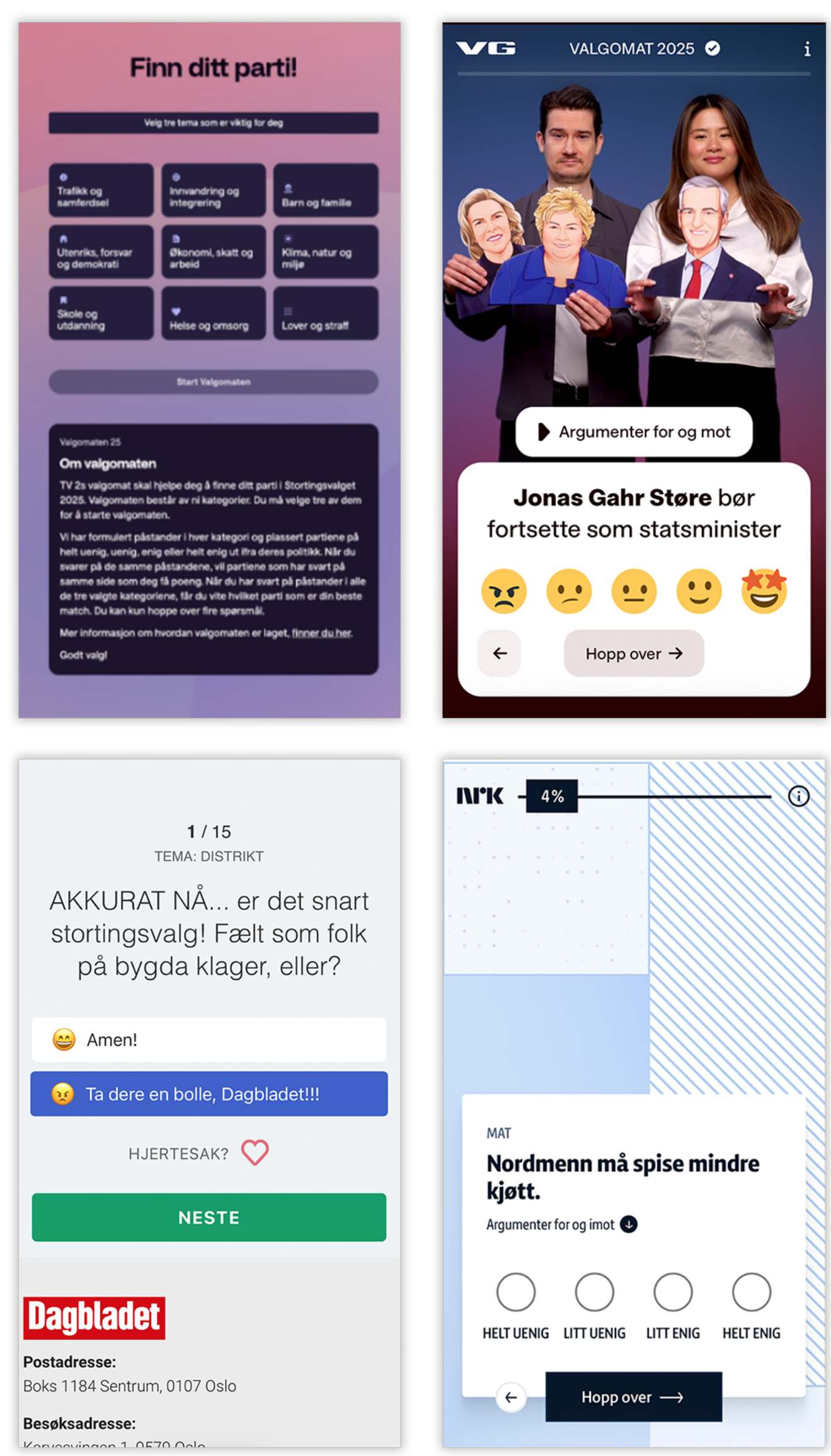

**Figure 2:** Screenshots of party selection quizzes from TV2 (t.l.), VG (t.r.), Dagbladet (b.l.), and NRK (b.r.).

# 5. Discussion

Two themes stand out from the analysis: First, the ambitious initial visions largely collapsed, indicating that the managers had unrealistic expectations towards the capacities of the technology and the pace of technological development. Furthermore, there was a conspicuous absence of concern about what we have called the internal threat: that use of GenAI could affect journalism in less noticeable ways, as some scholars have warned about (Shi and Sun 2024; Nandini et al. 2024).

In the planning stage at the start of the year, the AI managers aligned their strategies with prevailing sociotechnical imaginaries (Jasanoff, 2015) which frame testing the technology as part of an inevitable development path and that they could not afford to hesitate, despite uncertain consequences. This collective narrative seemed to legitimize editorial risk appetite related to GenAI, because not doing so could result in being "shot or eaten by someone else". These views echo prevailing sentiments in society about the future impact of AI, which are supported by narratives pushed by technology companies (Mager and Katzenbach 2021; Opdahl et al. 2023; Kotliar 2026).

While most of the ambitious initial ideas failed to materialize, the informants did appear to increase their uptake of GenAI, using the technology for tasks which previously were done without GenAI. This is consistent with past research which has shown that newsrooms primarily use AI for routine tasks rather than a great "AI revolution" which so far has not materialized (Opdahl et al. 2023; Amponsah and Atianashie 2024). A crucial question is whether the managers' grand visions of future GenAI may have distracted attention away from the threats associated with more mundane uses of GenAI – suffering from technological somnambulism, in Winner's (1988) terms.

There is a tension between this study's findings and the conclusions of Dodds and colleagues (2026), who found that the integration of GenAI in the practice of Dutch journalists was characterized by a "controlled change". Granted, the empirical data in this study to some degree echo the mechanisms of controlled change identified by Dodds and colleagues: Adaptive guidelines which balance GenAI use against ethical and professional norms, a practice of hands-on experimentation combined with human oversight, as well as an interest – to varying degrees – in assessing the systems critically. The fact that ambitious GenAI plans were shelved because the technology did not satisfy the organisations' standards for quality indicates that professional authority was given priority over the demand for technological innovation. All the newsrooms also had guidelines for AI use and offered training to their employees. However, our analysis also shows that such measures may not always be sufficient, as illustrated by the fact that one of the media organisations assumed that GenAI would simplify the task of making the party selection quiz so much that it could be done by an inexperienced student intern, resulting in mistakes slipping through.

Several of the managers emphasized that they trusted that they could rely on the critical instincts and competences of their journalists to guard against that GenAI-induced errors and undue influence, which reveals that the professional expertise of (human) journalists is a crucial factor in safeguarding against GenAI risks. As demonstrated by past research, GenAI may often appear convincing to human users, even when providing inaccurate information (Hackenburg et al. 2025; Salvi et al. 2025; Costello et al. 2026), and even highly skilled professionals are prone to overtrusting in LLMs (Klingbeil et al. 2024; Holbrook et al. 2024; Shekar et al. 2025). This raises the risk that LLMs may introduce problematic language into journalistic content, such as in the instance from TV2 when use of GenAI led to a question based on a politically charged concept. In this case, the error was caught by an experienced senior colleague.

If we consider the senior colleague's intervention as an instance of the safeguard provided by journalistic professional competence, this safeguard relied on at least three factors that might be threatened by GenAI integration. First, the senior colleague needed to have *time* to look through the quiz and give feedback. However, our findings as well as past research indicates that a main driver of AI integration is the idea that AI can increase efficiency (Moran and Shaikh 2022; Opdahl et al. 2023; Amponsah and Atianashie 2024; Simon 2024). If integration of GenAI is accompanied by demands that fewer journalists produce more content in less time, this may reduce the likelihood that journalists can spend time on this type of quality assurance. Second, the quiz maker had to *trust* the knowledge and routines of her more experienced colleague. However, the analysis revealed a factor that might challenge this trust: There was little transparency internally in the newsrooms regarding who used GenAI for what, and one of the quiz makers said that he was uncertain whether the political journalist who checked his quiz was also using GenAI to carry out the checks. This raises the possibility that newsrooms might inadvertently have GenAI systems do quality assurance on outputs from the same GenAI systems, without anyone being aware of it – leading to the risk of errors being reproduced or amplified. Finally, the third requirement for the safeguard is that the quiz maker needed to have *access to domain expertise within the newsroom staff*. If GenAI is increasingly used to automate journalistic tasks or replace editorial expertise, there is a risk that human journalists will lose the opportunity to develop the relevant skills.

Dodds and colleagues (2026) suggest that GenAI can only function as a complement to, and not a substitute for, human expertise. However, the analysis in this study shows that GenAI was indeed to some degree used as a substitute for human expertise, as the work of developing the party selection quizzes was delegated to inexperienced journalists and interns who might not have been given this job if it were not for GenAI. The quiz makers said that it was time-consuming to investigate facts in areas where they lacked expertise, and that it was also difficult to detect the errors that occurred in the GenAI tools. As noted in a seminal text on "Ironies of Automation" (Bainbridge 1983), automating tasks may lead to human skills deteriorating because they are not used regularly. This can over time lead to de-skilling of the whole newsroom. In this perspective, journalistic use of GenAI may lead to a Catch-22 situation: In order to be able to monitor the GenAI tools and prevent errors and undue influence, newsrooms rely on human expertise. But by using GenAI extensively, the newsrooms risk a deterioration of human expertise and de-skilling of the whole newsroom in the long term, preventing them from monitoring the GenAI systems adequately.

The findings in this study give reason to ask whether the newsrooms are too easily dismissing the risk that GenAI might subtly influence core values and journalistic priorities in election coverage. Respondents spoke of GenAI use as comparable to other mundane uses of technology such as Excel or phones, and rejected the possibility that GenAI had influenced their coverage, in spite of the fact that GenAI was routinely used to generate entire news stories, and shared examples of mistakes caused by GenAI use. Furthermore, there was little transparency about the newsrooms' use of GenAI, which runs contrary to recommendations for preventing undue influence from AI (Mahony and Chen 2025; de-Lima-Santos et al. 2025). In fact, transparency was further reduced ahead of the election campaign, as the media outlets stopped labelling content that had been produced with help from GenAI. This change may have been in part motivated by the concern, mentioned by one of the mangers, that AI labels lead readers to be more skeptical. However, as one of the journalists stated, lack of AI labelling might just as well cause skepticism towards all other content, since the readers now cannot know which has been generated by AI.

While it is not possible to determine in this study whether or not GenAI had undue or detrimental influence on the election coverage, the statements from participants raise a concern about whether journalists and editors are overly relaxed about this risk. Future research might combine newsroom observations and content analysis to investigate whether extensive use of GenAI leads to traceable changes or weighting in the news content itself. More research is also needed to assess whether increased use of GenAI and transparency (or lack of it) about such use affect audiences' trust in the news media.

## 6. Declaration of GenAI Use

The AI tool keenious.com was used in literature search for this study, but has not been used for generating any text appearing in the article. Interviews were transcribed using the transcription tool in Microsoft Word. Furthermore, an early draft of the article was translated from Norwegian to English using the current version of Microsoft Word for Mac (16.111.3). The translated text was subsequently proofread and extensively revised by the authors.

# 7. References

Amponsah, Peter N., and Atianashie Miracle Atianashie. 2024. “Navigating the New Frontier: A Comprehensive Review of AI in Journalism.” *Advances in Journalism and Communication* 12 (1): 1–17. https://doi.org/10.4236/ajc.2024.121001.

Bainbridge, Lisanne. 1983. “Ironies of Automation.” *Automatica* 19 (6): 775–79. https://doi.org/10.1016/0005-1098(83)90046-8.

Bender, Emily M., Timnit Gebru, Angelina McMillan-Major, and Shmargaret Shmitchell. 2021. “On the Dangers of Stochastic Parrots: Can Language Models Be Too Big? 🦜” *Proceedings of the 2021 ACM Conference on Fairness, Accountability, and Transparency* (New York, NY, USA), FAccT ’21, March 3, 610–23. https://doi.org/10.1145/3442188.3445922.

Bontridder, Noémi, and Yves Poullet. 2021. “The Role of Artificial Intelligence in Disinformation.” *Data & Policy* 3 (January): e32. https://doi.org/10.1017/dap.2021.20.

Bråten, Stein. 1983. *Dialogens vilkår i datasamfunnet: essays om modellmonopol og meningshorisont i organisasjons- og informasjonssammenheng*. Universitetsforlaget. https://www.nb.no/items/67ec5761a0f660b38ef282c632ba764c.

Costello, Thomas H., Kellin Pelrine, Matthew Kowal, et al. 2026. “Large Language Models Can Effectively Convince People to Believe Conspiracies.” arXiv:2601.05050. Preprint, arXiv, January 9. https://doi.org/10.48550/arXiv.2601.05050.

Dodds, Tomás, Wang Ngai Yeung, Claudia Mellado, and Mathias-Felipe De Lima-Santos. 2026. “On Controlled Change: Generative AI’s Impact on Professional Authority in Journalism.” *Journalism Studies* 27 (7): 1051–68. https://doi.org/10.1080/1461670X.2026.2616634.

Dodds, Tomás, Rodrigo Zamith, and Seth C. Lewis. 2026. “The AI Turn in Journalism: Disruption, Adaptation, and Democratic Futures.” *Journalism* 27 (3): 530–44. https://doi.org/10.1177/14648849251343518.

Dodge, Jesse, Maarten Sap, Ana Marasović, et al. 2021. “Documenting Large Webtext Corpora: A Case Study on the Colossal Clean Crawled Corpus.” In *Proceedings of the 2021 Conference on Empirical Methods in Natural Language Processing*, edited by Marie-Francine Moens, Xuanjing Huang, Lucia Specia, and Scott Wen-tau Yih. Association for Computational Linguistics. https://doi.org/10.18653/v1/2021.emnlp-main.98.

Fridman, M., R. Krøvel, and F. Palumbo. 2025. “How (Not to) Run an AI Project in Investigative Journalism.” *Journalism Practice* 19 (6): 1362–79. https://doi.org/10.1080/17512786.2023.2253797.

Gallegos, Isabel O., Ryan A. Rossi, Joe Barrow, et al. 2024. “Bias and Fairness in Large Language Models: A Survey.” *Computational Linguistics* 50 (3): 1097–179. https://doi.org/10.1162/coli_a_00524.

Gerring, John. 2007. “Is There a (Viable) Crucial-Case Method?” *Comparative Political Studies* 40 (3): 231–53. https://doi.org/10.1177/0010414006290784.

Hackenburg, Kobi, Ben M. Tappin, Luke Hewitt, et al. 2025. “The Levers of Political Persuasion with Conversational Artificial Intelligence.” *Science* 390 (6777): eaea3884. https://doi.org/10.1126/science.aea3884.

Holbrook, Colin, Daniel Holman, Joshua Clingo, and Alan R. Wagner. 2024. “Overtrust in AI Recommendations About Whether or Not to Kill: Evidence from Two Human-Robot Interaction Studies.” *Scientific Reports* 14 (1): 19751. https://doi.org/10.1038/s41598-024-69771-z.

Huang, Lei, Weijiang Yu, Weitao Ma, et al. 2025. “A Survey on Hallucination in Large Language Models: Principles, Taxonomy, Challenges, and Open Questions.” *ACM Trans. Inf. Syst.* 43 (2): 42:1-42:55. https://doi.org/10.1145/3703155.

Jasanoff, Sheila. 2015. “Future Imperfect: Science, Technology, and the Imaginations of Modernity.” In *Dreamscapes of Modernity: Sociotechnical Imaginaries and the Fabrication of Power*, edited by Sheila Jasanoff and Sang-Hyun Kim. University of Chicago Press. https://doi.org/10.7208/chicago/9780226276663.003.0001.

Jungherr, Andreas, Adrian Rauchfleisch, and Alexander Wuttke. 2026. “Artificial Intelligence in Election Campaigns: Perceptions, Penalties, and Implications.” *Political Communication* 43 (4): 585–606. https://doi.org/10.1080/10584609.2025.2611913.

Jungherr, Andreas, and Ralph Schroeder. 2023. “Artificial Intelligence and the Public Arena.” *Communication Theory* 33 (2–3): 164–73. https://doi.org/10.1093/ct/qtad006.

Kavtaradze, L., and B. Kalsnes. 2024. “AI-Powered Fact-Checking: Strategic Framing of AI Use.” *Strategic Communication–Contemporary Perspectives* 9177: 177–98.

Klein, Gary A., and Robert R. Hoffman. 1993. “Seeing the Invisible: Perceptual-Cognitive Aspects of Expertise.” In *Cognitive Science Foundations of Instruction*, 1st ed., edited by Mitchell Rabinowitz. Routledge. https://doi.org/10.4324/9781315044712-9.

Klingbeil, Artur, Cassandra Grützner, and Philipp Schreck. 2024. "Trust and Reliance on AI — An Experimental Study on the Extent and Costs of Overreliance on AI." *Computers in Human Behavior* 160 (November): 108352. https://doi.org/10.1016/j.chb.2024.108352.

Kotliar, Dan M. 2026. "Can't Stop the Hype: Scrutinizing AI's Realities." *Information, Communication & Society* 29 (3): 828–49. https://doi.org/10.1080/1369118X.2025.2531165.

Kreps, Sarah, and Doug Kriner. 2023. "How AI Threatens Democracy." *Journal of Democracy* 34 (4): 122–31.

Mager, Astrid, and Christian Katzenbach. 2021. "Future Imaginaries in the Making and Governing of Digital Technology: Multiple, Contested, Commodified." *New Media & Society* 23 (2): 223–36. https://doi.org/10.1177/1461444820929321.

Mahony, Simon, and Qing Chen. 2025. "Concerns about the Role of Artificial Intelligence in Journalism, and Media Manipulation." *Journalism* 26 (9): 1859–77. https://doi.org/10.1177/14648849241263293.

McIntosh, Timothy R., Tong Liu, Teo Susnjak, Paul Watters, Alex Ng, and Malka N. Halgamuge. 2024. "A Culturally Sensitive Test to Evaluate Nuanced GPT Hallucination." *IEEE Transactions on Artificial Intelligence* 5 (6): 2739–51. https://doi.org/10.1109/TAI.2023.3332837.

Møller, Lynge Asbjørn, Morten Skovsgaard, and Claes de Vreese. 2025. "Reinforce, Readjust, Reclaim: How Artificial Intelligence Impacts Journalism's Professional Claim." *Journalism* 26 (7): 1373–90. https://doi.org/10.1177/14648849241269300.

Moran, Rachel E., and Sonia Jawaid Shaikh. 2022. "Robots in the News and Newsrooms: Unpacking Meta-Journalistic Discourse on the Use of Artificial Intelligence in Journalism." *Digital Journalism* 10 (10): 1756–74. https://doi.org/10.1080/21670811.2022.2085129.

Nandini, Prabhat Indora, and R. K. Singh. 2024. "Artificial Intelligence in Journalism: An Overview of Its Applications and Uses." *Journal of Communication and Management* 3 (03): 237–42. https://doi.org/10.58966/JCM2024337.

Nemitz, Paul. 2018. "Constitutional Democracy and Technology in the Age of Artificial Intelligence." *Philosophical Transactions of the Royal Society A: Mathematical, Physical and Engineering Sciences* 376 (2133): 20180089. https://doi.org/10.1098/rsta.2018.0089.

Opdahl, Andreas L., Bjørnar Tessem, Duc-Tien Dang-Nguyen, et al. 2023. "Trustworthy Journalism through AI." *Data & Knowledge Engineering* 146 (July): 102182. https://doi.org/10.1016/j.datak.2023.102182.

Salvi, Francesco, Manoel Horta Ribeiro, Riccardo Gallotti, and Robert West. 2025. "On the Conversational Persuasiveness of GPT-4." *Nature Human Behaviour* 9 (8): 1645–53. https://doi.org/10.1038/s41562-025-02194-6.

Santos, Mathias-Felipe de-Lima-, Wang Ngai Yeung, and Tomás Dodds. 2025. "Guiding the Way: A Comprehensive Examination of AI Guidelines in Global Media." *AI & SOCIETY* 40 (4): 2585–603. https://doi.org/10.1007/s00146-024-01973-5.

Schia, Niels Nagelhus, Anne Sofie Molandsveen, Helle Sjøvaag, et al. 2025. *Artificial Intelligence and Democratic Elections – International Experiences and National Recommendations. Report by the Expert Group on Artificial Intelligence and Elections*. Ministry of Local Government and Regional Development. https://www.regjeringen.no/en/documents/artificial-intelligence-and-democratic-elections-international-experiences-and-national-recommendations-report-by-the-expert-group-on-artificial-intelligence-and-elections/id3086085/.

Shekar, Shruthi, Pat Pataranutaporn, Chethan Sarabu, Guillermo A. Cecchi, and Pattie Maes. 2025. "People Overtrust AI-Generated Medical Advice despite Low Accuracy." *NEJM AI*, ahead of print, May 13. World. https://doi.org/10.1056/AIoa2300015.

Shi, Yi, and Lin Sun. 2024. "How Generative AI Is Transforming Journalism: Development, Application and Ethics." *Journalism and Media* 5 (2): 582–94. https://doi.org/10.3390/journalmedia5020039.

Sikorski, Christian von, and Michael Hameleers. 2025. "Disinformation in the Age of Artificial Intelligence (AI): Implications for Journalism and Mass Communication." *Journalism & Mass Communication Quarterly* 102 (4): 941–57. https://doi.org/10.1177/10776990251375097.

Simon, Felix M. 2024. *Artificial Intelligence in the News: How AI Retools, Rationalizes, and Reshapes Journalism and the Public Arena*. Tow Center for Digital Journalism Publications. Tow Center for Digital Journalism, Columbia University. https://doi.org/10.7916/ncm5-3v06.

Sonni, Alem Febri, Hasdiyanto Hafied, Irwanto Irwanto, and Rido Latuheru. 2024. "Digital Newsroom Transformation: A Systematic Review of the Impact of Artificial Intelligence on Journalistic Practices, News Narratives, and Ethical Challenges." *Journalism and Media* 5 (4): 1554–70. https://doi.org/10.3390/journalmedia5040097.

Steensen, Steen, Bente Kalsnes, and Oscar Westlund. 2024. "The Limits of Live Fact-Checking: Epistemological Consequences of Introducing a Breaking News Logic to Political Fact-Checking." *New Media & Society* 26 (11): 6347–65. https://doi.org/10.1177/14614448231151436.

Steyvers, Mark, Heliodoro Tejeda, Aakriti Kumar, et al. 2025. "What Large Language Models Know and What People Think They Know." *Nature Machine Intelligence* 7 (2): 221–31. https://doi.org/10.1038/s42256-024-00976-7.

Stockwell, Sam, Megan Hughes, Phil Swatton, Albert Zhang, Jonathan Hall, and Kieran. 2024. *AI-Enabled Influence Operations: Safeguarding Future Elections*. Centre for Emerging Technology and Security, The Alan Turing Institute. https://cetas.turing.ac.uk/publications/ai-enabled-influence-operations-safeguarding-future-elections.

The Norwegian Media Authority. 2025. *Mediemangfaldsrekneskapen 2025. Mediemangfald i Eit Bruksperspektiv*. The Norwegian Media Authority. https://www.medietilsynet.no/globalassets/publikasjoner/mediemangfoldsregnskap/250505_bruksmangfaldsrapporten_2025.pdf.

VG. 2025. "VG endrer retningslinjene for bruk av kunstig intelligens." *Åpenhet - Redaksjonelle vurderinger i VG*, June 16. https://www.vg.no/informasjon/redaksjonelle-avgjorelser/iktzje.

Winner, Langdon. 1988. *The Whale and the Reactor: A Search for Limits in an Age of High Technology, Second Edition*. University of Chicago Press. https://press.uchicago.edu/ucp/books/book/chicago/W/bo49911830.html.

Zhang, Yue, Yafu Li, Leyang Cui, et al. 2025. "🧜Siren's Song in the AI Ocean: A Survey on Hallucination in Large Language Models." *Computational Linguistics* 51 (4): 1373–418. https://doi.org/10.1162/COLI.a.16.